\documentstyle[prl,aps,twocolumn]{revtex}
\input psfig
\begin{document}
\draft
 \newcommand{\mytitle}[1]{
 \twocolumn[\hsize\textwidth\columnwidth\hsize
 \csname@twocolumnfalse\endcsname #1 \vspace{1mm}]}
\mytitle{
\title{Kondo Correlations and Fano Effect in Closed 
AB-Interferometers}
\author{Walter Hofstetter,$^{1}$ J\"urgen K\"onig,$^{2,3}$ and 
Herbert Schoeller$^{4}$}
\address{
$^1$Theoretische Physik III, Elektronische Korrelationen und Magnetismus,
Universit\"at Augsburg, D-86135 Augsburg, Germany\\
$^2$Department of Physics, The University of Texas at Austin,
Austin, Texas 78712, USA\\
$^3$Institut f\"ur Theoretische Festk\"orperphysik, Universit\"at
Karlsruhe, D-76128 Karlsruhe, Germany\\
$^4$Theoretische Physik A, Technische Hochschule Aachen, D-52056 Aachen,
Germany}
\date{\today}
\maketitle
\begin{abstract}
We study the Fano-Kondo effect in a closed Aharonov-Bohm (AB) interferometer
which contains a single-level quantum dot and predict a frequency doubling of 
the AB oscillations as a signature of Kondo-correlated states.
Using Keldysh formalism, Friedel sum rule and Numerical Renormalization Group, 
we calculate the exact zero-temperature linear conductance $G$ as a function
of AB phase $\varphi$ and level position $\epsilon$. 
In the unitary limit, $G(\varphi)$ reaches its maximum $2e^2/h$ at 
$\varphi=\pi/2$.
We find a Fano-suppressed Kondo plateau for $G(\epsilon)$ similar to recent 
experiments.
\end{abstract}
\pacs{}
}
{\em Introduction.}
Recent measurements of transport through small semiconductor quantum dots
have revealed interesting quantum-coherence phenomena such as the Kondo effect 
\cite{kondo,hewson,kondo-theo,us} in quantum dots strongly coupled to leads, 
Aharonov-Bohm (AB) oscillations \cite{AB} of the current through 
multiply-connected geometries containing quantum dots, and Fano-type line shapes 
in multi-channel transport situations \cite{fano-exp}. 
In this letter, we study the combination of all these effects in an AB geometry
which contains a spin-degenerate single-level quantum dot, see 
Fig.~\ref{fig1}.

Interference between resonant transport through the quantum dot and the direct 
channel gives rise to asymmetric line shapes in the linear conductance as a 
function of gate or bias voltage, the well-known Fano effect \cite{fano-theo}. 
Such line shapes have been observed recently in linear transport through 
multi-level quantum dots \cite{fano-exp}, where the nature of the direct 
(``reference'') transmission path has not yet been fully clarified.  
Furthermore, scanning tunneling microscopy measurements of magnetic atoms on gold surfaces \cite{STM} 
yielded Fano lineshapes in the tunneling density of states, which have been 
successfully explained theoretically \cite{Zawadowski} under the assumption that 
only conduction electrons participate in tunneling.
The setup we propose (see Fig.~\ref{fig1}) has, however, the advantage that a 
controlled separate manipulation of both interfering paths is possible.
Enhanced AB oscillations due to the Kondo effect at low temperature were 
predicted \cite{Davidovich} for a similar geometry.

Here, we define two \emph{precise} criteria for detecting Kondo correlations 
in a closed geometry.
The first one is based on the interplay of Fano and Kondo physics.
We find that the Fano line shape in the Kondo regime is {\it qualitatively}
different as compared to a noninteracting system, where the Kondo effect is 
absent.
Furthermore, we find that the well-known Kondo plateau of increased 
conductance in the Coulomb-blockade regime is suppressed due to the 
Fano effect, and can even be inverted into a Kondo valley. 
We call this behaviour the Fano-Kondo effect.
The second criterion addresses the scattering phase.
In the Kondo regime, this phase is $\pi/2$ associated with unitary transmission
\cite{unitary}.
We predict that this can be detected in the AB phase $\varphi = 2\pi \Phi/\Phi_0$
dependence of the linear conductance $G(\varphi)$, which shows 
a \emph{frequency doubling} and a maximum at $\varphi=\pi/2$.
Here, $\Phi$ is the enclosed flux and $\Phi_0=hc/e$ denotes the flux quantum.
This is remarkable since it demonstrates that the symmetry relation 
$G(\varphi)=G(-\varphi)$ referred to as ``phase locking'', which is an exact 
property of (closed) two-terminal setups \cite{buettiker}, does not spoil the 
possibility to exhibit unitary scattering in the Kondo-correlated state.
Hence, it is not necessary to employ multi-terminal or open-geometry setups for
that task, which have been studied in order to avoid phase locking.
Their theoretical interpretation, however, relies on the assumption that 
multiple scattering events are absent, a condition which is difficult to control 
in experiments. 
This may be one reason for the discrepancy between recent phase shift 
measurements and theoretical predictions \cite{open}.
Another reason which favors the use of a closed-geometry setup from the 
practical point of view is the fact that the signal is strongly reduced in 
open geometries, since most of the emitted electrons are absorbed by terminals 
in the periphery.
\begin{figure}
\centerline{\psfig{figure=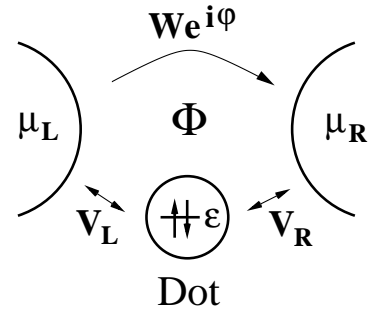,height=4cm}}
\vspace{.3cm}
\caption{AB interferometer with a quantum dot embedded.}
\label{fig1}
\end{figure}

{\em The Model.}
The model we consider is depicted in Fig.~\ref{fig1}.
Electrons driven through the device have either to go through the quantum dot, 
described by an Anderson model, or can be transfered directly.
The area enclosed by the two paths is penetrated by a magnetic flux $\Phi$.
The Hamiltonian $H=H_L+H_R+H_D+H_{T}+H_{LR}$ contains 
$H_r = \sum_{k \sigma} \epsilon_{kr} a^\dagger_{k\sigma r}a_{k\sigma r}$
for the left and right lead, $r=L/R$.
The isolated dot is described by 
$H_D = \epsilon \sum_\sigma n_\sigma + U n_\uparrow n_\downarrow$, where 
$n_\sigma = c^\dagger_\sigma c_\sigma$, and $\epsilon$ is the level energy 
measured from the Fermi energy of the leads.
Tunneling between dot and leads is modeled by $H_T = \sum_{k \sigma r} (V_r 
a^\dagger_{k\sigma r} c_\sigma + {\rm H.c.})$, where we neglect the energy 
dependence of the tunnel matrix elements $V_{L/R}$.
The intrinsic line width of the dot levels due to tunnel coupling to the leads
is (in the absence of the upper arm) $\Gamma=\Gamma_L+\Gamma_R$ with 
$\Gamma_{L/R}=2\pi|V_{L/R}|^2N_{L/R}$ where $N_{L/R}$ is the density of states 
in the leads.
The electron-electron interaction is accounted for by a charging energy 
$U=2E_C$ for double occupancy.
The transmission through the upper arm is modeled by the term
$H_{LR} = \sum_{kq\sigma} (W e^{i\varphi} a^\dagger_{k\sigma R} a_{q\sigma L} 
+ {\rm H.c.})$.
We choose a gauge in which the AB flux $\Phi$ enters via 
$\varphi = 2\pi \Phi/\Phi_0$ only the tunnel matrix element for the direct 
transmission.

{\em General current formula.}
The current from the right lead is given by 
$I = I_R = e {d \langle\hat n_R \rangle / dt}
= i(e/ \hbar) \langle [ \hat H, \hat n_R ] \rangle$.
The latter expression yields Green's functions which involve Fermi operators
for the leads and the dot,
\begin{eqnarray}
   I_R &=& -{e\over h} 
        \sum_{q\in L,k\in R,\sigma} \int d\omega \left[ 
        We^{i\varphi} G^<_{qk,\sigma}(\omega) + {\rm H.c.} \right]
\nonumber \\ 
        && - {e\over h}\sum_{k\in R,\sigma} \int d \omega 
        \left[ 
        V_R G^<_{dk,\sigma}(\omega) + {\rm H.c.} \right] 
        \, ,
\label{current_start}
\end{eqnarray}
with notations
$G^<_{qk,\sigma}(t) = i \langle a^\dagger_{k\sigma R} a_{q\sigma L}(t)\rangle$
and
$G^<_{dk,\sigma}(t) = i \langle a^\dagger_{k\sigma R} c_{\sigma}(t)\rangle$.
The indices $q$ and $k$ label the states in the left and right lead, 
respectively.
The index $d$ indicates that a dot electron operator is involved (in our 
simple model there is only one dot level). 
The first (second) line of Eq.~(\ref{current_start}) describes electron 
transfer from the left (from the quantum dot) to the right lead or vice versa.
This transfer can be a direct tunneling process or a complex trajectory 
through the entire device.

Our goal is to derive a relation between the current and Green's functions 
involving dot operators only.
To achieve this we employ the Keldysh technique for the Green's functions
$G^<$, $G^{\rm r}$, and $G^{\rm a}$, where $G^{\rm r}$ and $G^{\rm a}$ are the 
usual retarded and advanced Green's functions, respectively.
We write down Dyson-like equations, collect all contributions (including 
those with multiple excursions of the electrons to the leads and the quantum 
dot and arbitrary high winding number around the enclosed flux), and make use 
of the conservation of total current.
We get $I = (2e/ h) \int d\omega \, T(\omega) [f_L(\omega) - f_R(\omega)]$
with the total transmission probability $T(\omega)$ per spin given by
\begin{eqnarray}
   T(\omega) &=& T_{\rm b} 
        + \sqrt{\alpha T_{\rm b} R_{\rm b}}\cos{\varphi} \, \bar{\Gamma}\, 
        \mathrm{Re} \, G^r (\omega)
\nonumber \\ && 
        - {1\over 2} \left[ \alpha \left( 1-T_{\rm b}\cos^2\varphi \right) 
        -T_{\rm b} \right] \, \bar{\Gamma} \, \mathrm{Im} \, G^r (\omega)
\, , \label{general}
\end{eqnarray}
with $T_{\rm b}=4x/(1+x)^2$ being the background transmission probability, 
$R_{\rm b}=1-T_{\rm b}$, $x = \pi^2 W^2 N_L N_R$, and 
$\bar{\Gamma}=\Gamma/(1+x)$.
Asymmetry in the coupling of the dot level to the left and right lead is 
parametrized by $\alpha = 4\Gamma_L\Gamma_R/\Gamma^2$ (for resonant 
transmission the zero-temperature conductance through the quantum dot is
$G = \alpha e^2/h$ per spin).
The first term in Eq.~(\ref{general}) describes transmission through the 
upper arm.
The second and third term represent both transport through the quantum dot as
well as interference contributions.
We emphasize that the Green's function for the quantum dot entering 
Eq.~(\ref{general}) has to be evaluated {\em in the presence of the upper 
arm}.
The expansion of Eq.~(\ref{general}) up to first order in $\Gamma$ and $W$ has
been used \cite{koenig-gefen} to address the effect of spin-flip 
processes on the suppression of the interference signal. 
We note that Eq.~(\ref{general}) can be generalized to even more complex
scattering geometries within the formalism presented in 
Ref.~\cite{bruder-fazio-hs}.

Our result can be viewed as a generalization of the Landauer-B\"uttiker 
formula to the interacting case.
It includes interference effects and is especially suitable to describe
linear-response transport.
In this regime $T(\omega)$ is needed for zero bias voltage only, i.e., we can
calculate the {\em equilibrium} Green's function $G^{\rm r}$ by using the 
Numerical Renormalization Group (NRG) technique \cite{nrg} generalized to the 
case of two reservoir channels coupled to the dot \cite{com1}.
In the following, however, we focus on the $T=0$ case where, using Fermi liquid
properties, all relevant information can be extracted from the dot occupation 
number calculated by NRG.

{\em Friedel sum rule.}
For $U=0$, the Green's function can be calculated exactly. 
In equilibrium, the self-energy $\Sigma(\omega)$, defined by
$G^{\rm r}(\omega)=1 / [\omega - \epsilon - \Sigma(\omega)]$, reads
$\Sigma(\omega) = -(\bar{\Gamma}/2)\sqrt{\alpha}\sqrt{x}\cos{\varphi} - i\bar{\Gamma}/2$ and is
independent of $\omega$ for a flat band in the reservoirs.
In our calculations, this condition is satisfied due to $D \gg \Gamma$ where D 
is the half bandwidth (for a discussion of narrow-band effects see 
Ref.~\cite{narrow}).
For arbitrary $U$ we use the Friedel sum rule \cite{hewson} which 
at zero temperature yields $\mathrm{Im}\, \Sigma(0)= -\bar{\Gamma}/2$ and 
a renormalized level position $\epsilon+\mathrm{Re}\,\Sigma(0)$.
The latter is related to the resonant scattering phase shift 
$\delta_{\rm res}=\pi \langle n \rangle /2$ of the dot (where 
$\langle n \rangle = \sum_\sigma \langle n_\sigma\rangle$ is the total dot 
occupation) by
\begin{equation}
\label{friedel}
   e\equiv {2\over\bar{\Gamma}} \left[ \epsilon+\mathrm{Re}\,\Sigma(0) \right] 
        = \cot{\delta_{\rm res}}\, .
\end{equation}

With these properties we obtain the zero-temperature linear conductance 
$G = 2ge^2/h$ (or the dimensionless conductance $g=T(0)$) from 
Eq.~(\ref{general}). 
By some algebra, it can be recast into the the \emph{generalized Fano form}
\begin{equation}
\label{fano}
   g = 
T_{\rm b} {(e+q)^2\over e^2+1} + \alpha {\sin^2{\varphi}\over e^2+1}\, ,
\end{equation}
with the \emph{Fano parameter} 
\begin{equation}
\label{shift}
   -q=\cot{\delta}=\sqrt{\alpha R_{\rm b}/T_{\rm b}}\cos{\varphi}\,.
\end{equation}
Here, we have introduced the nonresonant phase shift $\delta$ which is due to 
scattering of free electrons at the weak link formed by the direct 
tunneling path. 
Eq.~(\ref{fano}) can then be expressed completely in terms of phase shifts: 
\begin{equation}
\label{phase-form}
   g = T_{\rm b} {\sin^2(\delta_{\rm res}-\delta)\over 
        \sin^2\delta}+ \alpha \sin^2\varphi\sin^2\delta_{\rm res}\,.
\end{equation}

\begin{figure}
\centerline{\psfig{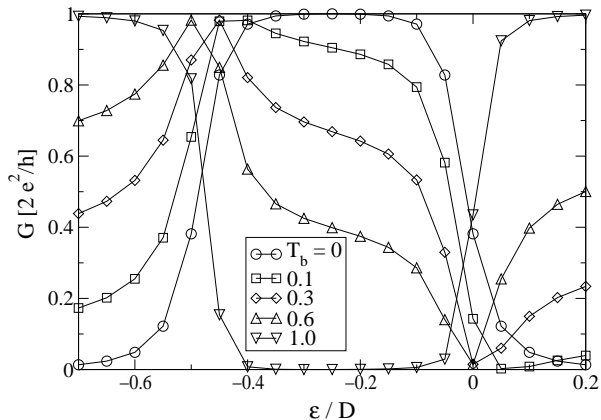}}
\vspace{.3cm}
\caption{Linear-response conductance at zero temperature 
as a function of level position $\epsilon$ 
for different values of the background transmission $T_{\rm b}$. 
We have used $U/D=0.5$ and $\Gamma/D=0.063$, where the half bandwidth
$D=1$ has been taken as the unit of energy. Note that due to the 
small ratio $\Gamma/D$, the results obtained here are 
cutoff--independent. The AB phase is $\varphi=0$. }
\label{fig2}
\end{figure}

{\em Discussion.}
Figure~\ref{fig2} shows the linear conductance as a function of level 
position $\epsilon$ for different values of $T_{\rm b}$ at vanishing AB phase
$\varphi=0$.
(For all figures we choose symmetric coupling, $\alpha = 1$.)
We find asymmetric line shapes of the peaks and dips around $\epsilon = 0$
and $\epsilon+U=0$, which indicates the presence of the Fano effect.

The Kondo effect shows up at $T_{\rm b}=0$ as a large plateau of unitary 
transmission.
At finite $T_{\rm b}$ the plateau survives but is reduced in height, similar to 
findings in recent experiments \cite{fano-exp}.
However, in these experiments the nature of the direct tunneling path has not 
yet been clarified. 
In particular, the direct tunneling strength cannot be tuned 
independently from the level broadening, which makes a quantitative 
comparison difficult.
It would, therefore, be highly desireable to realize the geometry shown in 
Fig.~\ref{fig1} in order to directly verify the reduction of the Kondo 
plateau. 
This reduction, which we call the Fano-Kondo effect, can be understood 
analytically from Eq.~(\ref{phase-form}). 
In the presence of interaction, the occupation of the dot is given 
by $\langle n \rangle \approx 1$ in the whole Coulomb blockade regime 
$-U+\bar{\Gamma}<\epsilon<-\bar{\Gamma}$, i.e., the resonant phase shift is 
$\delta_{\rm res}=\pi/2$, and Eq.~(\ref{phase-form}) simplifies to
\begin{equation}
\label{unitary}
   g_{\delta_{\rm res}=\pi/2} = \alpha \left( 1 - T_{\rm b}\cos^2\varphi
        \right) \, ,
\end{equation}
or $g = \alpha R_{\rm b}$ for $\varphi=0$ within the whole plateau. 
We emphasize that the Kondo effect does \emph{not} break down.
The reduction of the conductance plateau is rather due to interference between 
the lower (Kondo-) and the upper (reference) arm.
For noninteracting systems, the phase is $\pi/2$ at resonance ($\epsilon = 0$)
only, i.e., there is no plateau. 
The formation of a reduced \emph{Fano-Kondo plateau} is therefore a clear 
indication for Kondo correlations in the quantum dot under study.
Far away from the plateau, we get $\delta_{\rm res}\rightarrow 0,\pi$, and thus
$g = T_{\rm b}$ from Eq.~(\ref{phase-form}) as expected. 
For arbitrary $\delta_{\rm res}$ we get
\begin{equation}
\label{nophase}
   g_{\varphi=0} = \left[ \alpha+ (1-\alpha)T_{\rm b} \right]
        \sin^2(\delta_{\rm res} - \delta)\,.
\end{equation}
This means that the conductance has a peak (of height $2e^2/h$ for $\alpha=1$
or $T_{\rm b}=1$) at $\delta_{\rm res}=\delta\pm\pi/2$ and is zero for 
$\delta_{\rm res}=\delta$ in agreement with Fig.~\ref{fig2}. 
We emphasize the complementary behavior of the limiting cases of a closed
($T_{\rm b}=0$) and an open ($T_{\rm b}=1$) channel in the upper arm.
For $T_{\rm b}=0$ we obtain the usual Kondo plateau \cite{kondo,kondo-theo} 
but for $T_{\rm b}=1$ we have $\delta = \pi/2$ and {\em the conductance is 
zero in the Kondo valley but $2e^2/h$ outside}.

\begin{figure}
\centerline{\psfig{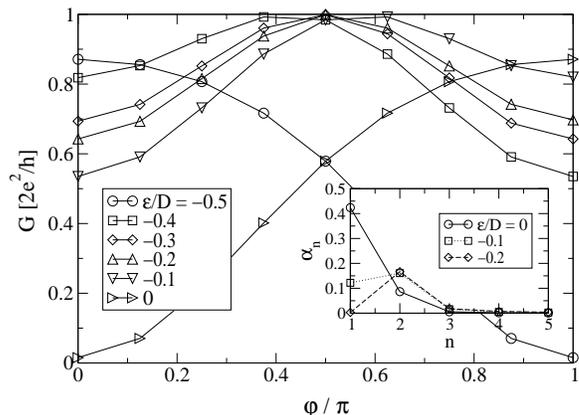}}
\caption{AB oscillations of the linear conductance for different level 
  positions $\epsilon$. 
  Parameters are identical to those in Fig.~\ref{fig2}, with a background 
  transmission of $T_b = 0.3$.
  Inset: absolute value of Fourier coefficients, $|\alpha_n|$. 
  In the Kondo regime, the first harmonic vanishes and $\alpha_2$ dominates,  
  thus effectively doubling the frequency of the AB oscillations.} 
\label{fig3}
\end{figure}

We now turn to the dependence of the conductance on the AB phase $\varphi$ and 
the question whether the phase $\delta_{\rm res}=\pi/2$ for unitary 
transmission in the Kondo regime can be detected from the AB oscillations
of the linear conductance for a {\em closed} AB interferometer. 
We have calculated the shape of the AB oscillations for different gate voltages 
in and outside the Kondo regime, as shown in Fig.~\ref{fig3}. 
Because of the phase locking property $G(\varphi) = G(-\varphi)$, we have
plotted only $\varphi \in [0,\pi]$. 
The inset shows the lowest coefficients of the Fourier expansion 
$G(\varphi) = \sum_n \alpha_n\, e^{i n \varphi}$.
Outside the Kondo regime ($\epsilon/D=0$ and $-0.5$), the AB oscillations are 
dominated by the lowest harmonic (period $2\pi$), with global extrema at 
$\varphi=0, \pi$. 
A drastic change occurs as soon as the dot is tuned into the Kondo regime: 
then, the AB oscillations show a maximum at $\varphi=\pi/2$ with the
universal conductance $2\alpha e^2/h$ {\em independent} of the value of the 
background transmission $T_{\rm b}$.
As a consequence (see inset of Fig.~\ref{fig3}), the $n=1$ Fourier coefficient 
vanishes, and the second harmonic dominates, thus effectively 
\emph{doubling the oscillation frequency}. 
Due to Kondo screening, this frequency doubling persists over a finite range 
of gate voltages corresponding to the conductance plateau. 
This effect, therefore, yields a conveniently measurable and precise criterion 
for Kondo correlations in a quantum dot.

\begin{figure}
\centerline{\psfig{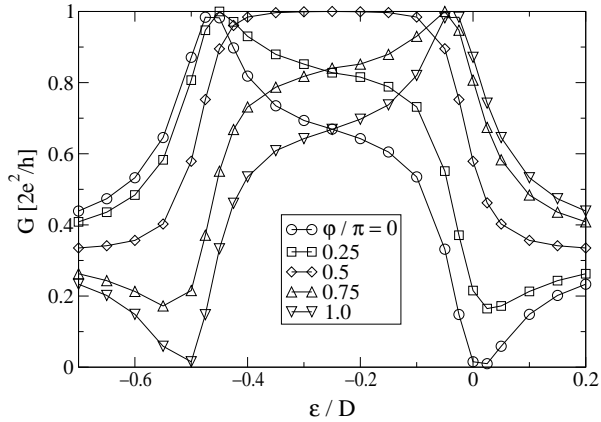}}
\vspace{.3cm}
\caption{Dependence of the Fano lineshape on the AB phase for the same 
parameters as in Fig.~\ref{fig2} and a background transmission of 
$T_{\rm b} = 0.3$.}
\label{fig4}
\end{figure}

Figure~\ref{fig4} illustrates the ``pinning'' of the AB maximum.
We show $G(\epsilon)$ for different AB phases $\varphi$. 
For $\varphi=\pi/2$ we recover a Kondo plateau with height 
\begin{equation}
\label{kondo}
   g_{\varphi=\pi/2}=\alpha + (T_{\rm b}-\alpha)\cos^2\delta_{\rm res}\,.
\end{equation}
A special case is $T_{\rm b}=\alpha$ where the conductance is given by
$2\alpha e^2/h$ for all level positions. 
Furthermore we note the property 
$G(\epsilon,\varphi)=G(-\epsilon-U,\varphi+\pi)$ which follows from 
particle-hole symmetry.

{\em Summary.}
We have studied the interplay of Kondo and Fano physics in the most basic model 
describing a closed AB interferometer which contains an interacting quantum dot 
in one of its arms.
We derived a current formula and used Friedel sum rule to calculate the exact 
linear conductance at zero temperature with the help of the Numerical 
Renormalization Group generalized to two-channel systems. 
We found a characteristic Fano-suppressed Kondo plateau.
Furthermore, we demonstrated that unitary transmission of a Kondo-correlated 
state can be identified by a frequency doubling of the AB oscillations and 
a conductance maximum at $\varphi=\pi/2$ with universal height.

{\em Acknowledgements.}
We would like to thank Y. Gefen, D. Vollhardt, and W.G. van der Wiel
for valuable discussions. This work is supported by the Deutsche 
Forschungsgemeinschaft under SFB 484 (W.H.) as well as under SFB 195 and the
Emmy-Noether program (J.K.).

{\em Note.}
While this paper was written up, a preprint \cite{bulka} was published,
in which a similar formula as Eq.~(\ref{general}) has been presented for the 
same model.
However, we believe that the result of Ref.~\cite{bulka} is incorrect since it
differs from our result by a factor $(1+x)$ in the second and third term.


\begin{references}
\bibitem{com1}
For vanishing AB phase $\varphi$ the system can be mapped 
on an effective single-channel model by taking appropriate linear combinations 
of the reservoir degrees of freedom.
\end{references}
\end{document}